\documentclass[10pt,a4paper]{amsart}
\usepackage[latin1]{inputenc}
\usepackage{amsmath}
\usepackage{amsfonts}
\usepackage{amssymb}
\usepackage{graphicx}
\usepackage{tikz} \usetikzlibrary{arrows,topaths,patterns,positioning}
\usepackage{amsthm}
\usepackage[margin=1.0in]{geometry}
\usepackage{cite}
\usepackage{pdflscape}

\renewcommand{\arraystretch}{1.5}

\newcommand{\Vir}{\mathfrak{Vir}}
\newcommand{\Heis}{\mathfrak{h}}
\newcommand{\Stag}{\mathcal{M}}
\newcommand{\Fock}{\mathcal{F}}
\newcommand{\FockS}{\mathcal{H}}
\newcommand{\univ}[1]{\mathcal{U}\left(#1\right)}
\newcommand{\ket}[1]{\left|#1\right>}
\newcommand{\bra}[1]{\left<#1\right|}
\newcommand{\matII}[4]{\left[\begin{array}{cc}
		#1	&	#2 \\
		#3	& 	#4 \\
	\end{array}\right]}
\newcommand{\de}{\mathrm{d}}

\newcommand{\normord}[1]{\mathop{:} #1 \mathop{:}}

\newtheorem*{theorem*}{Theorem}

\theoremstyle{definition}

\title{Free Field Realisations of Staggered Modules in 2D Logarithmic CFTs}
\author{Michael Cromer}
\address{Department of Theoretical Physics, Research School of Physics and Engineering (RSPE), and Mathematical Sciences Institute (MSI), Australian National University, Canberra, Australian Capital Territory, 2601}
\email{michael.cromer@anu.edu.au, mccromer.g@gmail.com}

\begin{document}
	
	\maketitle
	
	\begin{abstract}
		We utilise bosonic Fock spaces, considered as Virasoro modules, to make free field realisations of the so-called staggered modules of two-dimensional logarithmic conformal field theories. A general formula for the $\beta$-invariant of a staggered Fock module is derived, and found to agree with values previously known in the literature. In this way a large class of staggered modules is produced; one which provides an explicit free-field construction for many of those previously studied. We show how these modules can arise algebraically by including the modes of weight-$0$ fields into the algebra.
	\end{abstract}
	
	\tableofcontents
	
	\section{Introduction}
	
	Logarithmic theories are a relatively new topic of interest in the area of conformal field theory. They are required in certain 2D statistical models at phase transitions, notably in critical percolation and dilute polymers (e.g. \cite{DJS10, VJS12, Ri09}), as well as in more abstract settings such as the fusion ring of highest weight Virasoro modules \cite{CrRi13}. Their presence is indicated by logarithmic terms appearing in certain field correlators \cite{Gur93, Gur13}. At the representation-theoretic level, this requires the presence of generalised eigenvectors of the Virasoro $L_0$ mode. Despite much activity, logarithmic theories are still less well understood than their non-logarithmic cousins. Their representation theory, being that of indecomposable representations of some vertex operator algebra (the so-called \textit{staggered modules}) rather than of highest weight modules, is correspondingly richer and more complicated \cite{Roh96, KyRi09}.
	
	To date, several key features of the simplest types of logarithmic Virasoso representations have been identified. Roughly speaking, these representations are constructed by `gluing' two well-understood Virasoro representations (i.e., highest weight modules) together indecomposably. This is achieved by nontrivially extending one module by the other, in the manner of a non-split short exact sequence. The key features include universal invariants associated with the representations themselves, which appear in physically relevant computations such as correlation functions. A main object of interest is the $b$- or $\beta$-\textit{invariant}\footnote{Also called \textit{anomaly numbers}, \textit{logarithmic couplings}, and \textit{indecomposability parameters} by various other authors.}, seen in the field theory as a coefficient in an OPE and in the representation theory as an inner product between certain vectors \cite{DJS10, VJS11, VJS12}. The value of $\beta$ is not necessarily the same in the two cases, instead being related by a possible combinatorial factor \cite{MaRi08}. Two highest weight spaces may possess many different indecomposable `gluings', resulting in a range of possible $\beta$ invariants. It has been found, however, that the value of $\beta$ uniquely fixes the staggered module up to isomorphism \cite{KyRi09}.
	
	The fusion product of two $\Vir$ representations is a kind of modified product module. Ordinarily, the tensor product of two representations at central charge $c$ would result in a representation at charge $c+c$. The fusion product of these spaces, while still possessing the product space as its base, instead has $\Vir$ action altered so that the central charge of the resulting representation is also $c$. In this way we build up an algebra structure on the $\Vir$ representations at a fixed central charge. From here, for instance, we might study the smallest such algebra which contains the minimal models of some rational highest weight theory. Interestingly, such fusion algebras do not generally close on minimal models, or even on highest weight modules. In some cases, it has been necessary to include staggered modules \cite{Gab03}. 
	
	Here we stress the difference between genuine highest weight modules and graded modules which merely contain a highest weight vector. The latter contains the former as a subset, which consists of those modules which not only contain a highest weight vector, but are entirely generated by it. It is now expected that the relevant $\Vir$ modules for logarithmic theories are in general \textit{not} highest weight. The name \textit{Virasoro Kac module} has been used for those modules whose characters arise in the continuum scaling limit of logarithmic minimal models. There is already strong evidence that free field realisations, or Feigin-Fuchs modules, will be necessary to explain the appearance of Virasoro Kac modules in logarithmic theories \cite{Ras11, DRR15}.
	
	Only particular staggered modules arise in applications, possessing certain $\beta$ invariants. However, these quantities remain difficult to compute in general. In this paper we suggest a new method of constructing staggered modules, making use of bosonic Fock spaces as the underlying representations. The extra structure afforded us by the free boson allows us to derive an explicit expression for the $\beta$-invariant for all staggered Fock modules of a particular type. When we compare the values of $\beta$ predicted by this expression to those already known in the literature, we find in every case an exact match. Since this identifies the staggered modules as isomorphic, we may call this a free-field realisation of these staggered modules.
	
	Section~\ref{sec::background} contains a short background of staggered modules, as well as a discussion of the submodule embedding structures of Fock spaces as Virasoro modules. The subject of vertex operators is also introduced, particularly their use in constructing explicit module homomorphisms of Fock spaces. This is done in some detail, recovering an explicit algebraic form for these operators, with a view to their use in the construction of staggered Fock modules in later sections. For this purpose we also introduce a method for generating a family of operators with particular commutation relations from a single module homomorphism.
	
	In Section~\ref{sec:StaggeredFockModules}, we define staggered Fock modules. We construct a particular family of these modules using the vertex operators of Section \ref{sec::background}. Our main result is then an explicit expression for the $\beta$-invariant for these staggered modules. After this, we examine in detail which of these modules appear in the context of the reducible (rational) $(p,q)$ theories. We also show how to interpret $\Delta h = 0$ staggered Fock modules through an induced representation construction, and how this may also be extended to $\Delta h \neq 0$ staggered Fock modules.
	
	Finally, in Section~\ref{sec::conclusion}, we summarise the results obtained. We offer some discussion about future applications.
	
	\section{Background and Notation}
	\label{sec::background}
	
	\subsection{logCFT and Staggered Virasoro Modules}
	\label{subsec::StaggeredVirMods}
	
	For our purposes, a (\textit{rank $2$}) \textit{staggered module} $\Stag$ is any $\Vir$-module such that there is a non-split short exact sequence
	\begin{equation}
	\label{eqn:SES}
	0 \longrightarrow X_L \overset{\iota}{\longrightarrow} \Stag \overset{\pi}{\longrightarrow} X_R \longrightarrow 0	
	\end{equation}
	where $X_L$ and $X_R$ (called the \textit{left} and \textit{right} module respectively) are $\Vir$ highest weight modules and the maps $\iota, \pi$ module homomorphisms. We also demand that the $\Vir$ zero mode $L_0$ act non-diagonalisably on $\Stag$. Let $x_i$ be the highest weight vector of $X_i$, and $h_i$ its $L_0$ eigenvalue. Write $v_L = \iota(x_L)$ and choose some representative $v_R \in \pi^{-1}(\{x_R\})$.
	
	$\Stag$ may still be given a grading by splitting $\Stag$ into commuting diagonal ($L_0^d$) and nilpotent ($L_0^n$) parts, then grading $\Stag$ into $L_0^d$ eigenspaces. Equivalently we can say that an element $v \in \Stag$ is of fixed grade if the grades of both $\iota^{-1}(v)$ and $\pi(v)$ agree (in their respective spaces). It may be desired, in practical applications, to only consider $\Vir$ representations on $\Stag$ which respect this grading\footnote{We shall briefly discuss examples of representations which do \textit{not} in Section~\ref{sec:StaggeredFockModules}.}.
	
	We have, as usual,
	\[
	L_0 v_L = h_L v_L
	\]
	but crucially
	\[
	L_0 v_R = h_R v_R + w \qquad w \in \iota(X_L).
	\]
	To follow with established terminology, we say that $v_R$ is the logarithmic partner of $v_L$.
	
	When the $\Vir$ representation respects the grading on $\Stag$ it is not too hard to show that the difference in vacuum conformal weights, $\Delta h  = h_L - h_R$, must be an integer in order for $\Stag$ to exist at all, and that $\iota^{-1}\left((L_0 - h_R)v_R\right) = \iota^{-1}(w)$ must be a singular vector of $X_L$. Its conformal weight must then be $h_R = h_L - \Delta h$. $\Delta h$ must in fact be negative in order for $\Stag$ to be a staggered module in the sense we desire -- it is possible to set $\Delta h \geq 0$, but the resulting module is of a different type (e.g. reducible Verma modules are one example of this).
	
	The representative $v_R \in \pi^{-1}(x_R) \subset \Stag$ is only defined up to addition of a vector $u \in X_L$ of $L_0$ eigenvalue $h_R$ (called a \textit{gauge transform} by some authors). We must therefore turn to quantities independent of this choice in order to study the essential properties of $\Stag$. One such quantity is called the \textit{$\beta$-parameter} or \textit{$\beta$-invariant}. It only depends upon the chosen normalisation of the singular vector $(L_0 - h_R)v_R$ and characterises the isomorphism class of $\Stag$ \cite{KyRi09}. 
	
	Let $U_{\Delta h} \in \univ{\Vir}$ be a creation operator such that $U_{\Delta h}$ generates the singular vector of grade $h_R$ in the Verma cover of $X_L$. We fix the normalisation\footnote{The relevant coefficient is nonzero; see \cite{Ast95}.} of both $U_{\Delta h}$ and of $v_R$ so that 
	\[
	U_{\Delta h} = L_{-1}^{\Delta h} + \cdots
	\]  
	and
	\[
	(L_0 - h_R)v_R = U_{\Delta h} v_L.
	\]
	The benefit of this normalisation is that it does not depend on the choice of PBW basis ordering,
	
	Then we define
	\begin{equation}
	\label{eqn:betaDefn}
	\beta(\Stag) := \left< v_L, (U_{\Delta h})^\dagger v_R \right>.
	\end{equation}
	where the adjoint $^\dagger$ is the anti-involution which sends $L_{n}$ to $L_{-n}$, $c$ to $c$. Note that since $(U_{\Delta h})^\dagger v_R$ is an element of $\iota(X_L)$, the inner product used here is the Shapovalov form of this submodule of $\Stag$, as inherited from $X_L$.
	
	\subsection{Fock Space as a $\Vir$-Module}
	\label{subsec::FockSpaceAsAVirModule}
	
	Fock spaces are highest weight modules of the infinite dimensional Heisenberg algebra $\Heis$. This algebra has basis
	\[
	\{\alpha_n | n \in \mathbb{Z}\} \cup \{\mathbf{1}\}
	\]
	with relations
	\[
	[\alpha_m, \alpha_n] = m\delta_{m, -n} \mathbf{1}, \qquad [\alpha_n, \mathbf{1}] = 0.
	\]
	A Fock space is then the $\Heis$-module induced from a $1$-dimensional representation of the Borel subalgebra $\Heis_{\geq 0}$ with basis $\{\alpha_n | n \geq 0\}\cup \{\mathbf{1}\}$ where the positive-indexed modes act as multiplication by $0$, $\mathbf{1}$ as multiplication by $1$, and $\alpha_0$ as multiplication by a constant $\eta \in \mathbb{C}$. We call the spanning element of this representation the \textit{vacuum} vector with \textit{vacuum} (or \textit{momentum})  \textit{eigenvalue} $\eta$, denoted by $\ket{\eta}$. The Fock space induced from this vacuum is denoted by $\Fock_\eta$:
	\[
		\Fock_\eta = \univ{\Heis} \otimes_{\Heis_{\geq 0}} \ket{\eta}.
	\]
	
	The basis elements $\alpha_{n}$ are also the modes of the weight-$1$ primary field
	\[
	\partial a(z) = \sum_{n\in \mathbb{Z}} \alpha_n z^{-n-1}
	\]
	and from this $\Fock_\eta$ may be made into a graded $\Vir$ module. We write
	\[
	T(z) = \sum_{n \in \mathbb{Z}} L_n z^{-n-2}
	\]
	for the energy-momentum tensor, whose modes are the basis elements of the Virasoro algebra. Then by equating
	\[
		T(z) := \frac{1}{2}\normord{\partial a(z)^2} + \lambda \partial^2 a(z),
	\]
	for some choice of constant $\lambda \in \mathbb{C}$, we obtain the following action of $L_n$ on $\Fock_\eta$:
	\[
		L_n = \frac{1}{2}\sum_{k \in \mathbb{Z}} \normord{\alpha_k \alpha_{n-k}} - \lambda (n+1) \alpha_n
	\]
	(the central charge $c$ is parametrised by $\lambda$ as $c = 1 - 12\lambda^2$). The vacuum vector also has $L_0$ eigenvalue
	\begin{equation}
	\label{eqn:h(eta,lambda)}
	h_\eta = \frac{1}{2}\eta(\eta - 2\lambda).
	\end{equation}
	The space $\Fock_\eta$ with this particular $\Vir$ representation is called $\Fock_{\eta, \lambda}$. When $c$ is fixed, we often abbreviate this to $\Fock_\eta$ as $\lambda$ is assumed understood.
	
	When comparing two Fock spaces $\Fock_\eta$ and $ \Fock_{\eta + \xi}$ at the same central charge, an often-useful quantity is the difference in vacuum conformal weight. We keep the notation $\Delta h$ for this quantity, which for Fock spaces is equal to 
	\begin{equation}
	\label{eqn:deltah}
	\Delta h = h_{\eta + \xi} - h_\eta = (\eta - \lambda)\xi + \frac{1}{2}\xi^2.
	\end{equation}
	
	Given the very simple commutation relations of $\Heis$, $\Fock_\eta$ is always irreducible as an $\Heis$ module. Its structure as a $\Vir$ module is not always as simple. The reducibility and decomposability of Fock space Virasoro modules is well-studied (e.g. \cite{FeFu90}), and for $c < 1$ there is a neat classification scheme arising from that of reducible highest-weight Virasoro modules in terms of coprime positive integers. We refer of course to the well-known Kac table of Verma modules, a convenient way to display the zeroes of the Kac determinant formula (a good introduction to the subject is in \cite{dFMS}).
	
	For the so-called reducible $(p, q)$ models, then, where $p, q$ are coprime positive integers (arbitrarily take the convention $q > p$), we have the following:
	\begin{align}
	\begin{split}
	\label{eqn:hrs}
	c_{p,q} 		&= 1 - \frac{6(q-p)^2}{pq} \\
	h_{r, s} &= \frac{(rq - sp)^2 - (q - p)^2}{4pq}, \quad r, s \in \mathbb{Z}^+
	\end{split}
	\end{align}
	where $r, s \in \mathbb{Z}^+$ are a second pair of positive integers (not necessarily coprime). Given fixed $p, q$, all such $r, s$ result in reducible Fock spaces which share a common central charge. Not all $r, s$ result in distinct modules, due to the periodicity of $h_{r,s}$. The so-called minimal models occur at $0 < r < p$, $0 < s < q$, and it is these entries which comprise the $(p, q)$ Kac table. In studying logarithmic theories we find it necessary to work with arbitrary $r, s > 0$, and the correspondingly extended table is called (unsurprisingly) the \textit{extended} $(p, q)$ Kac table. 
	
	In any given extended table only the first $p$ many rows have distinct values, the rows being vertically periodic. The entries shift to the right by $q$ many places each period. The entire table is symmetric when reflected about any one $(r, s) = (np, np)$ entry. In what follows, entries such that $(r,s)=(mp,np)$ will be called \textit{corner} entries. We also adopt the name \textit{edge} entries to refer to those for which only one of $r = mp$ or $s = nq$ holds. All others we refer to as \textit{bulk} entries. The `type' of submodule embedding structure of a Verma module -- and of Fock spaces with the same $c$ and $h$ -- only depends on whether it is a corner, edge or bulk entry in the table \cite{IoKo, FeFu90}.
	
	\begin{figure}[]
		\[
		\begin{array}{|c|ccccccc}
		\hline r\backslash s& 1 &2 &3 &4 &5 & 6 &\cdots\\ \hline
			1 & 0& 0& \frac{1}{3} & 1& 2& \frac{10}{3} & \cdots \\
			2 & \frac{5}{8}& \frac{1}{8}& \frac{-1}{24}&\frac{1}{8} & \frac{5}{8}& \frac{35}{24} & \\
			3 & 2& 1& \frac{1}{3} &0 &0 & \frac{1}{3} & \\
			4 & \frac{33}{8}& \frac{21}{8} & \frac{35}{24} &\frac{5}{8} & \frac{1}{8}& \frac{-1}{24}& \cdots \\
			\vdots & \vdots& & & & &\vdots & \\
		 \end{array}
		\]
		\caption{ The first few entries of the extended $(2, 3)$ Kac table. The minimal models of the (non-extended) table occupy the top-left corner. In this particular example there are two such entries, and both have $h=0$. \label{fig:pqTableExample}}
	\end{figure}
	
	The entries of the extended tables also correspond to reducible Fock modules. We find from $c = 1 - 12\lambda^2$ and (\ref{eqn:h(eta,lambda)}) that setting
	\begin{align}
	\begin{split}
	\label{eqn:etars+/-}
	\lambda 			&= \lambda_{p,q}^\pm = \pm\frac{(q - p)}{\sqrt{2pq}} \\
	\eta = \eta_{r,s}^\pm  	&= \frac{1}{\sqrt{2pq}}\left((q - p) \pm (rq - sp)\right).
	\end{split}
	\end{align}
	defines a reducible Fock space whose central charge and $\Vir$ highest weight match those of the $(r,s)$ entry of the extended $(p,q)$ Kac table. There are thus two distinct Fock spaces for any given $p, q, r, s$. They share the same vacuum conformal weight as well as their central charge. Here $\eta_{r, s}^\pm$ assumes the choice $\lambda = \lambda^+$, but changing to $\lambda^-$ merely has the effect of swapping the two representations (there is a sign change, but it is irrelevant since $\lambda^\pm = -\lambda^\mp$ introduces a complementary one):
	\[
	\lambda^+ \leftrightarrow \lambda^- \quad \implies \quad \eta_{r, s}^+ \leftrightarrow -\eta_{r,s}^-
	\]
	Due to the extreme similarities between the Fock space and the Verma cases, we will often find it convenient to refer to `the' extended $(p, q)$ Kac table, not bothering to distinguish between the two. We will make this difference clear, when necessary. Also, when discussing the reducible $(r,s)$ entries of a particular extended $(p,q)$ Kac table, we will write $\Fock_{(r, s)}^\pm$ instead of $\Fock_{\eta^\pm_{r,s}}$. 
	
\begin{figure}
		\centering
		\[
		\underbrace{\hspace{-0.9cm}\begin{tikzpicture}[>=stealth',semithick,auto]
		\tikzstyle{v3} = [circle, minimum width=8pt, fill, inner sep=0pt]
		\tikzstyle{v1}  = [circle, minimum width=8pt, draw, inner sep=0pt]
		\tikzstyle{v2}   = [circle, minimum width=8pt, draw, inner sep=0pt, path picture={\draw (path picture bounding box.south east) -- (path picture bounding box.north west) (path picture bounding box.south west) -- (path picture bounding box.north east);}]
		\node[v3] (0) {};
		\node[] (label0) [left = 0.1cm of 0.north] {$\ket{\eta_{r,s}^+}$};
		\node[v3] (1l) [below left of = 0, yshift = 0.1cm] {} 
		edge [<-] (0);
		\node[] (labelrs) [left = 0.01cm of 1l] {$(rs)$};
		\node[v1] (1r) [below right of = 0, yshift = -0.33cm] {}
		edge [->] (0);
		\node[v2] (2l) [below of = 1l] {}
		edge [<-] (1r)
		edge [->] (1l);
		\node[v2] (2r) [below of = 1r] {}
		edge [<-] (1r)
		edge [->] (1l);
		\node[v3] (3l) [below of = 2l] {}
		edge [<-] (2r)
		edge [<-] (2l);
		\node[v1] (3r) [below of = 2r] {}
		edge [->] (2r)
		edge [->] (2l);
		\node[v2] (4l) [below of = 3l] {}
		edge [<-] (3r)
		edge [->] (3l);
		\node[v2] (4r) [below of = 3r] {}
		edge [<-] (3r)
		edge [->] (3l);
		\node (etcl) [below of = 4l] {$\vdots$}
		edge [<-] (4r)
		edge [->] (4l);
		\node (etcr) [below of = 4r] {$\vdots$}
		edge [<-] (4r)
		edge [->] (4l);
		\end{tikzpicture} \quad
		\begin{tikzpicture}[>=stealth',semithick,auto]
		\tikzstyle{v3} = [circle, minimum width=8pt, fill, inner sep=0pt]
		\tikzstyle{v1}  = [circle, minimum width=8pt, draw, inner sep=0pt]
		\tikzstyle{v2}   = [circle, minimum width=8pt, draw, inner sep=0pt, path picture={\draw (path picture bounding box.south east) -- (path picture bounding box.north west) (path picture bounding box.south west) -- (path picture bounding box.north east);}]
		\node[v3] (0) {};
		\node[] (label0) [left = 0.1cm of 0.north] {$\ket{\eta_{r,s}^-}$};
		\node[v1] (1l) [below left of = 0, yshift = 0.1cm] {} 
		edge [->] (0);
		\node[] (labelrs) [left = 0.01cm of 1l] {$(rs)$};
		\node[v3] (1r) [below right of = 0, yshift = -0.33cm] {}
		edge [<-] (0);
		\node[v2] (2l) [below of = 1l] {}
		edge [->] (1r)
		edge [<-] (1l);
		\node[v2] (2r) [below of = 1r] {}
		edge [->] (1r)
		edge [<-] (1l);
		\node[v1] (3l) [below of = 2l] {}
		edge [->] (2r)
		edge [->] (2l);
		\node[v3] (3r) [below of = 2r] {}
		edge [<-] (2r)
		edge [<-] (2l);
		\node[v2] (4l) [below of = 3l] {}
		edge [->] (3r)
		edge [<-] (3l);
		\node[v2] (4r) [below of = 3r] {}
		edge [->] (3r)
		edge [<-] (3l);
		\node (etcl) [below of = 4l] {$\vdots$}
		edge [->] (4r)
		edge [<-] (4l);
		\node (etcr) [below of = 4r] {$\vdots$}
		edge [->] (4r)
		edge [<-] (4l);
		\end{tikzpicture}}_{\underset{\text{``Braid type''}}{\text{Bulk}}} \qquad \qquad
		\underbrace{\hspace{-1cm}\begin{tikzpicture}[>=stealth',semithick,auto]
		\tikzstyle{v2} = [circle, minimum width=8pt, fill, inner sep=0pt]
		\tikzstyle{v1}  = [circle, minimum width=8pt, draw, inner sep=0pt]
		\node[v2] (0) {};
		\node[] (label0) [left = 0.1cm of 0.north] {$\ket{\eta_{r,s}^+}$};
		\node[v2] (1) [below of = 0] {} 
		edge [<-] (0);
		\node[] (labelrs) [left = 0.01cm of 1] {$(rs)$};
		\node[v1] (2) [below of = 1] {}
		edge [->] (1);
		\node[v2] (3) [below of = 2] {} 
		edge [<-] (2);
		\node[v1] (4) [below of = 3] {}
		edge [->] (3);
		\node (etc) [below of = 4] {$\vdots$}
		edge [->] (4);   
		\end{tikzpicture} \quad
		\begin{tikzpicture}[>=stealth',semithick,auto]
		\tikzstyle{v2} = [circle, minimum width=8pt, fill, inner sep=0pt]
		\tikzstyle{v1}  = [circle, minimum width=8pt, draw, inner sep=0pt]
		\node[v2] (0) {};
		\node[] (label0) [left = 0.1cm of 0.north] {$\ket{\eta_{r,s}^-}$};
		\node[v1] (1) [below of = 0] {} 
		edge [->] (0);
		\node[] (labelrs) [left = 0.01cm of 1] {$(rs)$};
		\node[v2] (2) [below of = 1] {}
		edge [<-] (1);
		\node[v1] (3) [below of = 2] {} 
		edge [->] (2);
		\node[v2] (4) [below of = 3] {}
		edge [<-] (3);
		\node (etc) [below of = 4] {$\vdots$}
		edge [<-] (4);      
		\end{tikzpicture}}_{\underset{\text{``Chain type''}}{\text{Edge}}} \qquad \qquad
		\underbrace{\hspace{-1cm}\begin{tikzpicture}[>=stealth',semithick,auto]
		\tikzstyle{v2} = [circle, minimum width=8pt, fill, inner sep=0pt]
		\node[v2] (0) {};
		\node[] (label0) [left = 0.1cm of 0.north] {$\ket{\eta_{r,s}^\pm}$};
		\node[v2] (1) [below of = 0] {};
		\node[] (labelrs) [left = 0.01cm of 1] {$(rs)$};
		\node[v2] (2) [below of = 1] {};
		\node[v2] (3) [below of = 2] {};
		\node[v2] (4) [below of = 3] {};
		\node (etc) [below of = 4] {$\vdots$};   
		\end{tikzpicture}}_{\underset{\text{``Point type''}}{\text{Corner}}}
		\]\vspace{-20pt}\caption{The various possible submodule embedding structures of $\Fock_{r,s}^\pm$. Conformal weight increases down the page, with arrows indicating nontrivial acti on of the Virasoro algebra. Filled nodes denote singular vectors, unfilled denote subsingular ones which become singular in the quotient by the maximal proper submodule. Partially filled nodes (seen in the bulk modules) denote subsingular vectors which are in a sense `intermediate' as they belong to the maximal proper submodule itself, and are in fact subsingular within it. Note how the choice of $\eta_{r, s}^\pm$ changes the embedding structure -- in particular the role of the vector at grade $rs$, marked on the diagrams. It should be noted that in the bulk the first proper singular vector of $\Fock_{r,s}^+$ is not always of lower conformal weight than that of $\Fock_{r,s}^-$ (which is at grade $(p-r)(q-s)$ for the non-extended table entries, for instance); this apparently being the case is simply due to our choice of diagram. \label{fig:FockEmbedding}}
	\end{figure}
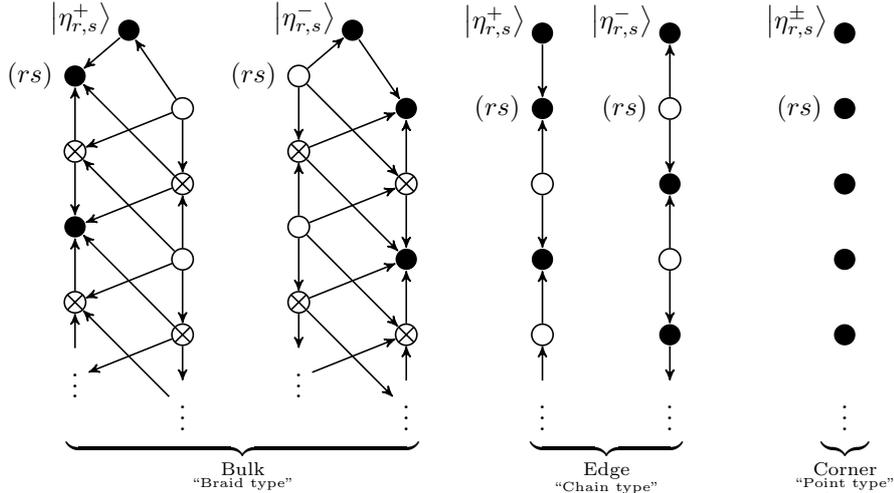

	The singular vectors of $\Fock_{r, s}^\pm$ occur at conformal weights where singular vectors appear in the corresponding Verma module, but this relationship does not hold in reverse. Instead, the unpaired singular vectors of the Verma module correspond to \textit{subsingular} vectors of the Fock space. Subsingular vectors are vectors which are not singular themselves, but which become singular in the quotient module of $\Fock_{r,s}^\pm$ by an appropriate submodule (usually the maximal proper submodule, but not always). Just as with Verma modules, there are three broad types of submodule embedding structure, each corresponding to corner, edge, or bulk entries. There is a further sub-categorisation depending on the choice of $\eta_{r,s}^\pm$ (Figure~\ref{fig:FockEmbedding}).
	
	We would like to point out that the left and right modules in a particular sequence (\ref{eqn:SES}) need not belong to an extended $(p,q)$ Kac table, or even be reducible at all. This is also the case for the staggered Fock spaces we construct in Section~\ref{sec:StaggeredFockModules}. While the extended Kac tables are indeed extremely pertinent to physical applications, we should stress that from a purely representation-theoretic point of view our results are relevant at arbitrary central charges $c$.
	
	\subsection{Vertex Operators}
	
	The field $\partial a(z)$ introduced in Section~\ref{subsec::FockSpaceAsAVirModule} is the derivative of the weight-$0$ field
	\[
	a(z) = q + \ln(z)\alpha_0 - \sum_{n \neq 0}\frac{1}{n}\alpha_n z^{-n}.
	\]
	The mode $q$ is a new operator whose commutation relations are
	\[
	[\alpha_n, q] = \delta_{n,0} \mathbf{1}.
	\]
	Due to this, it is useful in constructing maps between Fock spaces. We have:
	\[
		\alpha_0 e^{\mu q} \ket{\eta} = (\eta + \mu) e^{\mu q} \ket{\eta}
	\]
	so we may identify $e^{\mu q} \ket{\eta}$ with $\ket{\eta+\mu}$, and $e^{\mu q}: \Fock_\eta \to \Fock_{\eta+\mu}$ is a homomorphism of Fock spaces as $\Heis$ modules (but not, we note, as $\Vir$ modules).
	
	We therefore define the vertex operator of momentum $\mu$ to be
	\begin{align*}
	V_\mu(z) 	&= \normord{\exp(\mu a(z))} \\	&= e^{\mu q } z^{\mu \alpha_0} \exp\left(\mu \sum_{n > 0} \frac{\alpha_{-n}}{n}z^n\right) \exp\left(-\mu \sum_{n>0} \frac{\alpha_n}{n}z^{-n}\right).
	\end{align*}
	Whenever $V_\mu(z)$ has trivial monodromy about $0$,
	\[
		V_\mu(z) = \sum_{n \in \mathbb{Z}} V_n z^{-n-h_\mu}
	\]
	is well-defined, and its zero mode
	\[
		\oint_{z = 0} z^{h_\mu - 1}V_\mu(z) \de z = V_0 : \Fock_\eta \to \Fock_{\eta+\mu}
	\]
	is a $\Vir$-intertwiner between Fock spaces provided that $\lambda = \frac{1}{2}\mu - \frac{1}{\mu}$ (i.e., $h_\mu = 1$). Within a fixed $(p, q)$ model, this gives two possible solutions:
	\begin{equation}
	\label{eqn:mu+/-}
		\mu^+ = \sqrt{\frac{2q}{p}}, \qquad \mu^- = -\sqrt{\frac{2p}{q}}.
	\end{equation}
	(We stress that this is the $q$ of Section~\ref{subsec::FockSpaceAsAVirModule}, not the mode of $a(z)$ introduced above! In the sequel, we will attempt to indicate which $q$ is intended when context fails to make this clear.) As expected,
	\[
	\lambda^+ \leftrightarrow \lambda^- \quad \implies \quad \mu^+ \leftrightarrow -\mu^-.
	\]
	Also at fixed $(p, q)$, we find
	\[
	\eta_{r, s}^\pm + \mu^+ = \eta_{r\pm2, s}^\pm \qquad \eta_{r, s}^\pm + \mu^- = \eta^\pm_{r, s\pm 2},
	\]
	so vertex operators preserve the choice $(\pm)$ of vacuum $\eta_{r, s}^\pm$, with $\mu^+$ changing the $r$ index and $\mu^-$ changing the $s$, both in steps of $2$ with an appropriate sign. Note that the central charge $c$ is never changed. 
	
	Intertwiners from single vertex operators are useful, but are applicable in relatively few instances. More general intertwiners can be constructed via the normal ordering several vertex operators. For the normal ordering of $N$ many vertex operators, we have 
	\begin{align}
	\begin{split}
	\label{eqn:NormOrdNMany}
	\normord{V_\mu(z_1) \cdots V_\mu(z_N)} &= e^{N\mu q} \prod_{i<j}(z_i - z_j)^{\mu^2}\prod_i z_i^{\mu\alpha_0} \\ &\quad\times\exp\left(\mu \sum_{n > 0} \frac{\alpha_{-n}}{n}p_n\right)\exp\left(-\mu \sum_{n > 0} \frac{\alpha_{n}}{-n}p_{-n}\right)
	\end{split}
	\end{align}
	where $p_n = (z_1^n + \cdots z_N^n)$ is the $n$th power sum. When the context is clear, we abbreviate (\ref{eqn:NormOrdNMany}) as $V_{N\times \mu}(z)$.
	
	The `zero mode' $V_0$ is now harder to extract, but provided that $\lambda = \frac{1}{2}\mu - \frac{1}{\mu}$ and $\eta = \lambda - \frac{1}{2}N \mu - M \frac{1}{\mu}$ for some integer $M$, then there exists a contour $\Gamma$ such that 
	\begin{equation}
	\label{eqn:V_0fromNFields}
	\oint_\Gamma V_{N\times \mu}(z) \de z_1 \cdots \de z_N = V_0: \Fock_\eta \to \Fock_{\eta + N\mu}
	\end{equation}
	is a \textit{bona fide} non-trivial intertwiner \cite{TK86}.
	
	The first of these two conditions is the familiar requirement that the vertex operators be Virasoro primaries of conformal weight $1$. The second, the monodromy condition, is actually clearer when re-expressed in terms of the vacuum conformal weights of the domain and image spaces:
	\[
	\Delta h = N\left(\mu \eta + \frac{1}{2}(N-1)\mu^2 + 1\right),
	\]
	so is equivalent to the requirement that $\Delta h$ be an integer. Using these restrictions, we see there may only exist at most a single pair of modules related by any one given triple $(N, \Delta h, \mu)$. Indeed, if $V_0: \Fock_{\eta_1} \to \Fock_{\eta_2}$, is nontrivial, then
	\begin{equation}
	\eta_1 = \frac{1}{2}(1-N)\mu + \frac{\Delta h - N}{N\mu}, \qquad \eta_2 = \frac{1}{2}(1+N)\mu + \frac{\Delta h - N}{N\mu}
	\end{equation}
	and so
	\begin{align}
	\begin{split}
	&h_{\eta_1} = \frac{1}{2}\left[\frac{1}{4}(N^2 - 1)\mu^2 + (1 - \Delta h) + \frac{\Delta h^2 - N^2}{N^2 \mu^2} \right], \\
	&h_{\eta_2} = \frac{1}{2}\left[\frac{1}{4}(N^2 - 1)\mu^2 + (1 + \Delta h) + \frac{\Delta h^2 - N^2}{N^2 \mu^2} \right].
	\end{split}
	\end{align}
	The central charge is also determined by $c = 1 - 12(\frac{\mu}{2} - \frac{1}{\mu})^2$, uniquely determining the two modules.
	
	\section{Staggered Fock Modules}
	\label{sec:StaggeredFockModules}
	
	We study staggered modules for which the left and right modules are Fock spaces. In order to satisfy the requirement that they also be highest weight modules, we may restrict to any highest weight submodule containing the vacuum. Indeed, for a Fock space $\Fock_{\eta, \lambda}$, let $\FockS_{\eta, \lambda}$ denote the submodule generated by the action of the Virasoro algebra on only the vacuum $\ket{\eta}$:
	\[
	\FockS_{\eta, \lambda} := \univ{\Vir}\cdot \ket{\eta} \subseteq \Fock_{\eta, \lambda}
	\]
	We may then study when it is possible to construct a staggered module
	\[
	0 \longrightarrow X_L = \FockS_{\eta_L, \lambda} \longrightarrow \Stag \longrightarrow \FockS_{\eta_R, \lambda} = X_R \longrightarrow 0.
	\]
	
	However, we note that it is in principle not necessary to demand that the left and right modules be highest weight spaces, given that this has in the past often been a requirement because the modules under consideration already satisfied this condition. Fock spaces are in general \textit{not} highest weight spaces, so we will relax this condition. If desired, one easily sees that quotients may be taken to return to the modules $\FockS$ without disrupting the staggered structure.
	
	For what follows, then, we define a (\textit{rank 2}) \textit{staggered Fock module} to be any module $\Stag$ which fits into the non-split short exact sequence
	\begin{equation}
	\label{eqn:FockSES}
		0 \longrightarrow \Fock_L \overset{\iota}{\longrightarrow} \Stag \overset{\pi}{\longrightarrow}\Fock_R \longrightarrow 0
	\end{equation}	
	such that the maps $\iota, \pi$ are $\Vir$ module homomorphisms, $L_0$ is non-diagonalisable on $\Stag$, and $\Fock_L$ and $\Fock_R$ are Fock spaces with a common central charge $c$. We do not require $\iota, \pi$ to be $\Heis$ module homomorphisms (obviously if they were, they could be extended to $\Vir$ homomorphisms $\tilde{\iota}, \tilde{\pi}$, but the converse \textit{restriction} is not necessarily possible. It might be fruitful to examine which conditions allow us to infer $\Heis$ module homomorphisms from $\Vir$ ones in a staggered Fock module).
	
	
	In order to analyse the structure of staggered Fock modules, we must look first at the algebraic constraints required by the existence of a $\Vir$-representation. Write the action of the $\Vir$ generators on $\Stag \cong_{\text{Vec}} \Fock_L \oplus \Fock_R$ (in the obvious basis) as
	\[
	L_n = \matII{L_n}{V_n}{0}{L_n}\: \forall n \in \mathbb{Z}, \qquad c = \matII{c}{0}{0}{c}
	\]
	where the $V_n$ are a family of operators $\Fock_R \to \Fock_L$. For this to be a genuine $\Vir$-module, the off-diagonal terms $V_n$ must satisfy
	\begin{equation}
	\label{eqn:StaggeredFockRelation}
	[L_m, V_n] - [L_n, V_m] = (m-n)V_{m+n}.
	\end{equation}
	This is certainly the case if the $V_n$ are the modes of some $\Vir$-primary field of dimension $1$, although we note that this is not a necessary condition. In the sequel, we will refer to any such collection of maps $\{V_n | n \in \mathbb{Z}\}$ as \textit{staggering operators}. Our notation for staggering operators has been suggestively chosen so as to match with the modes of the vertex operators. Indeed, there is no obstruction to constructing a staggered Fock module with $V_\mu(z)$ giving the staggering operators, provided the series expansion in modes $V_n$ is well-defined.
	
	It seems `obvious' in this case to pair each $L_n$ with the corresponding vertex operator mode $V_n$, as then the action of the Virasoro algebra on $\Stag$ respects the grading discussed in Section~\ref{subsec::StaggeredVirMods}. However, there of course exist $\Vir$ representations on $\Stag$ which do not. One such representation is produced by shifting the index of the staggering operator associated with each $L_n$ by some constant $a \in \mathbb{Z}$, so that 
	\[
	L_n = \matII{L_n}{V_{n+a}}{0}{L_n}
	\]
	on $\Stag$. Then these index-shifted staggering operators also satisfy (\ref{eqn:StaggeredFockRelation}). As we are interested in graded representations only, we will ignore cases like this. s
	
	It is not clear that it is possible to construct a staggered module from the $N$-fold product $V_{N\times \mu}(z)$ of vertex operators. Such a construction would require a consistent definition of the non-zero modes of $V_{N\times \mu}(z)$, a potentially difficult task. Instead, we make use of the Jacobi identity to \textit{define} a family of staggering operators which will always satisfy (\ref{eqn:StaggeredFockRelation}). Suppose $\phi_0: \Fock_{\eta_1} \to \Fock_{\eta_2}$ is an intertwiner of Fock spaces considered as $\Vir$-modules, i.e. $[L_m, \phi_0] = 0$ for all $m$. Then define
	\[
	\phi_n := \frac{1}{k_\phi}[\alpha_n, \phi_0]
	\]
	where $k_\phi = \eta_2 - \eta_1$ is the necessary constant such that 
	\[
	\phi_0 = \frac{1}{k_\phi} [\alpha_0, \phi_0]
	\] 
	holds true.
	
	Then the family $\{\phi_n | n \in \mathbb{Z}\}$ satisfies (\ref{eqn:StaggeredFockRelation}), and forms a valid set of staggering operators. In particular, we consider the case $\phi_0 = V_0$ for $V_0$ arising from the composition of $N$ many vertex operators $V_\mu(z)$. Then
	\begin{equation}
	\label{eqn:VnDef}
	V_n := \frac{1}{N\mu}[\alpha_n, V_0].
	\end{equation}
	We note that this normalisation $k_{V_0} = N\mu$ also appears at the field-theoretic level, when taking the commutator of fields $[\partial a(z), V_{N\times\mu}(w)]$.
	
	One advantage of the above construction is that the $\beta$-invariants for these staggered modules will be manifestly insensitive to overall scalings of $V_0$. Most importantly, given a family of non-trivial staggering operators constructed this way, the staggered Fock module is \textit{guaranteed} to exist. This is because our construction is explicit and depends only on the intertwiner $V_0$. In a reducible $(p, q)$ model, this means exactly two inequivalent staggered Fock modules for each choice of $(N, \Delta h)$, provided $V_{N \times \mu}$ is well-defined -- one for each of $\mu^\pm$.
	
	However, to make explicit calculations possible, we must determine the exact form of $V_0$. We do so via some polynomial generating function identities \cite{Macdonald}. We have:
	\[
	\exp\left(\sum_{k>0} x_k z^k\right) = \sum_{n\in \mathbb{Z}} s_n(x) z^n
	\]
	where
	\[
	s_n(x) = \sum_{\lambda_1 + 2\lambda_2 + \cdots + n\lambda_n = n} \left(\prod_{i=1}^n \frac{x_i^{\lambda_i}}{\lambda_i!}\right)
	\]
	is the $n$th \textit{elementary Schur polynomial} \cite{KaRa}. Obviously $s_{n}(x) = 0$ for $n<0$, but we keep the sum over $\mathbb{Z}$ as it simplifies later notation.
	
	In order to make use of this generating function, we projectivise $V_{N \times \mu}(z)$ by the change of variables
	\begin{equation}
	\label{eqn:ProjectiveVariables}
	z_1 \mapsto x, \qquad z_{i} \mapsto x y_i \quad (i \neq 1).
	\end{equation}
	We note that the standard CFT conditions $z_i \neq 0, z_j, \infty$ become here $x \neq 0, \infty$ and $y_i \neq 0, 1, y_j, \infty$, suggesting that the contour $\Gamma$ should be chosen as a loop in $x$ about the origin followed by a (generalised) Pochhammer contour in the $y_i$.
	
	Write 
	\[
	P_n(y) = s_n \left(\frac{\mu}{k} \alpha_{-k} (1 + p_k(y))\right), \quad  Q_n(y) = s_n \left(\frac{-\mu}{k} \alpha_{k} (1 + p_k(y^{-1})) \right)
	\]
	for the elementary Schur polynomials in the creation and annihilation Heisenberg modes respectively, once projectivised (recall $p_k$ is the $k$th power sum, and that Heisenberg creators and annihilators commute amongst themselves, so ordering within each $P_n$ and $Q_n$ is unimportant). We substitute these expressions into (\ref{eqn:V_0fromNFields}) and, in the new variables, the integral to extract the intertwiner becomes
	\begin{align}
	\begin{split}
	V_0 &= e^{N\mu q}\oint_\Gamma \sum_{n \in \mathbb{Z}} \left(\sum_{k\in \mathbb{Z}} P_k Q_{k+n}\right)x^{-n-1 + \Delta h} \prod_{i < j} (y_i - y_j)^{\mu^2} \prod_i (1- y_i)^{\mu^2} y_i^{\mu \eta_R} \de y_i \de x \\
	&=e^{N\mu q}\oint_{x = 0}\oint_\mathcal{S} \sum_{n \in \mathbb{Z}} \Omega_n x^{-n-1 + \Delta h} \prod_{i < j} (y_i - y_j)^{\mu^2} \prod_i (1- y_i)^{\mu^2} y_i^{\mu \eta_R} \de y_i \de x
	\end{split}
	\end{align}
	with $\Omega_n = \sum_{k\in\mathbb{Z}} P_kQ_{k+n}$, and $\mathcal{S}$ the contour for the \textit{generalised Selberg integral} in $N-1$ variables \cite{War08}.
	
	It can be shown that if the integral does not vanish, it results in a nontrivial intertwiner $V_0$ \cite{TK86}. We evaluate the integral in $x$ separately. Up to scalings (which we safely ignore), this means
	\begin{equation}
	\label{eqn:V0Def}
	V_0 = e^{N\mu q}\oint_\mathcal{S} \Omega_{\Delta h} \prod_{i < j} (y_i - y_j)^{\mu^2} \prod_i (1- y_i)^{\mu^2} y_i^{\mu \eta_R} \de y_i.
	\end{equation}
	Evaluation of $V_0 \ket{\eta_R}$ is particularly simple since
	\[
	e^{N\mu q}\Omega_{\Delta h} \ket{\eta_R} = P_{\Delta h}(y) \ket{\eta_L},
	\]
	so we get a sum over partitions $\tau$ of $(-\Delta h)$:
	\begin{equation}
	\label{eqn:V0Evaluation}
	V_0 \ket{\eta_R} = \left(\sum_{|\tau| + \Delta h = 0} \frac{1}{\prod_{i = 1}^{n}n_\tau(i)!}B^{(\tau)}_N(\mu) \prod_{\tau_i \in \tau} \frac{\mu \alpha_{-\tau_i}}{\tau_i}\right)\ket{\eta_L}.
	\end{equation}
	Here $n_\tau(i)$ counts the number of times $i$ appears in $\tau$, and $B^{(\tau)}_N(\mu)$ is the evaluation of
	\begin{equation}
	\label{eqn:BmuIntegralFunction}
	B^{(\tau)}_N(\mu) = \oint_\mathcal{S} \prod_{\tau_j \in \tau} \left(1 + p_{\tau_j}(y)\right)\prod_{i < j} (y_i - y_j)^{\mu^2} \prod_i (1- y_i)^{\mu^2} y_i^{\mu\eta_R} \de y_i,
	\end{equation}
	still an admittedly complicated expression. We will see how we are able to avoid its computation altogether. In fact, we may even use the structure of the staggered module and its singular vectors to \textit{evaluate} integrals of the above type, if so desired (c.f. Appendix~\ref{app:Selberg}).
	
		\subsection{Admissible $\Fock_L$ and $\Fock_R$}
	\label{subsec:AdmissibleLandR}
	
	Here we follow through with the above comments about the uniqueness of the staggered Fock module with a given set of staggering operators. In particular we identify the exact pairs of entries in the extended $(p, q)$ Kac tables which correspond to a given choice of $(N, \Delta h)$. Suppose that $\eta_R = \eta_{r, s}^\pm$. Given a fixed $\Delta h$, we solve $\eta_L = \eta_R + N\mu$ for $(r, s)$. We find four separate relations, each corresponding to a different choice of $\eta_R = \eta_{r,s}^\pm$ and $\mu = \mu^\pm$ (Figure~\ref{fig:rsRelation}).
	
	\begin{figure}[h!]
		\[
		\begin{array}{|c|cc|}
		\hline s	&	\eta_{r,s}^+	&	\eta_{r,s}^- \\ \hline
		\mathbf{\mu^+}		&	(r+N)\frac{q}{p} -\frac{\Delta h}{N} 	&	(r - N)\frac{q}{p} + \frac{\Delta h}{N} \\
		\mathbf{\mu^-}		&	(r+\frac{\Delta h}{N})\frac{q}{p} -N		&	(r-\frac{\Delta h}{N}) \frac{q}{p} + N \\ \hline 
		\end{array}
		\]\caption{$s$ given $r$ for $\Fock_R = \Fock_{r,s}^\pm$ within a staggered Fock module.\label{fig:rsRelation}}
	\end{figure}
	
	One may easily check that the left and right modules are insensitive to the choice of $\eta^\pm$, so two distinct solutions can be retrieved by fixing either $\eta_R = \eta^+_{r,s}$ or $\eta_R = \eta_{r,s}^-$ and then using the two different $\mu^\pm$ (this is why we have chosen to present $\beta$ in terms of $\mu$ in Section~\ref{subsubsec:TheoremAndExamples}).
	
	A linear relation with reduced rational slope $\frac{q}{p}$ and reduced rational intercept $\frac{b}{a}$ possesses integral solutions if and only if $a$ divides $p$. If it possesses one such solution $(r_0, s_0)$, then necessarily infinitely many integral solutions $(r_0 + k p, s_0+k q)$ exist, $k \in \mathbb{Z}$. By examining the table in Figure~\ref{fig:rsRelation}, we see that integral solutions always exist when $N = 1$. Indeed, for solutions to exist for arbitrary $N \in \mathbb{Z}^+$, we require
	\[
	N | p\Delta h  \quad(\mu^+), \qquad N | q\Delta h  \quad(\mu^-)
	\]
	since these are the conditions under which the denominator of the intercept divides the denominator of the gradient for the rational linear relations in Figure~\ref{fig:rsRelation}.
	
	
	Let us suppose for the moment that $\frac{\Delta h}{N}$ is an integer. When this is the case, the two distinct staggered modules have short exact sequences:
	\[
	0\longrightarrow \Fock_{(np-N,nq +\frac{\Delta h}{N})}^{(-)} \longrightarrow \Stag \longrightarrow \Fock_{(np+N,nq + \frac{\Delta h}{N})}^{(-)} \longrightarrow 0 \qquad (\mu^+)
	\]
	and
	\[
	0\longrightarrow \Fock_{(np+\frac{\Delta h}{N},nq - N)}^{(-)} \longrightarrow \Stag \longrightarrow \Fock_{(np+\frac{\Delta h}{N},nq + N)}^{(-)} \longrightarrow 0 \qquad (\mu^-)
	\]
	The left and right modules, and thus the staggered modules themselves, are independent of the choice of $n\in\mathbb{Z}$. Recall that the superscript $(-)$ is used to indicate $\eta = \eta^-$. Choosing $\eta^+$ instead does not give distinct $\Stag$, but does relate entries at different $(r, s)$ in the extended table. The sequences then are:
	\[
	0\longrightarrow \Fock_{(np+N,nq-\frac{\Delta h}{N})}^{(+)} \longrightarrow \Stag \longrightarrow \Fock_{(np-N,nq-\frac{\Delta h}{N})}^{(+)} \longrightarrow 0 \qquad (\mu^+)
	\]
	and
	\[
	0\longrightarrow \Fock_{(np-\frac{\Delta h}{N},nq+N)}^{(+)} \longrightarrow \Stag \longrightarrow \Fock_{(np-\frac{\Delta h}{N},nq-N)}^{(+)} \longrightarrow 0 \qquad (\mu^-)
	\]
	Again, all choices of $n$ are identical. Staggered modules in these four sequences are isomorphic if and only if they share the same $\mu$.
	
	These particular solutions occur in rectangular patterns symmetrically arranged around corner entries of the extended $(p, q)$ table with $(r, s) = (np, nq)$. Left modules occur on the principal diagonal corners of this pattern, right modules on the off-diagonal. $\mu^+$ maps vertically and $\mu^-$ horizontally. Sequences involving $\eta^+$ have $\Fock_L$ below or to the right of $\Fock_R$ in the table, and those involving $\eta^-$ have $\Fock_L$ above or to the left. Figure~\ref{fig:ExtendedKacExample} gives pictorial examples of this.
	
	A similar pattern appears when $\frac{\Delta h}{N} \notin \mathbb{Z}$. If we assume the conditions for the existence of a solution are satisfied, then locating viable $(r, s)$ in the table is tantamount to solving for $k, l \in \mathbb{Z}$ such that
	\begin{equation}
	\label{eqn:klLiinearRelations}
	kq - lp = \frac{p\Delta h}{N} \in \mathbb{Z} \quad (\mu^+), \qquad kq - lp = -\frac{q\Delta h}{N} \in \mathbb{Z} \quad (\mu^-),
	\end{equation}
	possible since $p$ and $q$ are coprime. Then the sequences become (again identical for all $n$):
	\begin{align}
	&\left.\begin{array}{c}
	0\longrightarrow \Fock_{(np+ k+N,nq+l)}^{(+)} \longrightarrow \Stag \longrightarrow \Fock_{(np+ k - N,nq+l)}^{(+)} \longrightarrow 0\\
	0\longrightarrow \Fock_{(np+k -N,nq-l)}^{(-)} \longrightarrow \Stag \longrightarrow \Fock_{(np+k +N,nq-l)}^{(-)} \longrightarrow 0\end{array}\qquad\right\} \: \mu^+ \\
	&\left.\begin{array}{c}
	0\longrightarrow \Fock_{(np+k,nq+l+N)}^{(+)} \longrightarrow \Stag \longrightarrow \Fock_{(np+k,nq+l-N)}^{(+)} \longrightarrow 0\\
	0\longrightarrow \Fock_{(np-k,nq+l-N)}^{(-)} \longrightarrow \Stag \longrightarrow \Fock_{(np-k,nq+l+N)}^{(-)} \longrightarrow 0\end{array}\qquad\right\} \: \mu^-,
	\end{align}
	though we note that the particular $(k, l)$ need not be the same across the two cases $\mu^\pm$.
	
	These modules exhibit the same symmetric rectangular distribution around particular points in the extended $(p,q)$ table, but not around the `principal diagonal' corner entries $(np, nq)$. Instead they are located symmetrically around the $(np+k,nq)$ and $(np, nq+l)$ entries (for $\mu^+$ and $\mu^-$ respectively). A cursory examination of the relations (\ref{eqn:klLiinearRelations}) shows that, necessarily, $p|k$ ($\mu^+$) and $q|l$ ($\mu^-$). Therefore these more general sequences appear symmetrically around \textit{other} corner entries, at $(r, s) = (mp, nq)$ with $m\neq n$. The other properties mentioned (direction of mappings in the table, relative locations of left and right modules, etc.) remain the same.
	
	\begin{figure}
		\centering
		\tikzset{>=latex}
		\begin{tikzpicture}
		\draw[step=1cm,gray,very thin] (-2.9,-2.9) grid (3.9,3.9);
		\draw[pattern=north west lines, pattern color=black] (0.0,-2.9) rectangle (1.0,3.9);
		\draw[pattern=north east lines, pattern color=black] (-2.9,0.0) rectangle (3.9,1);
		
		\node at (1.5, 2.5) (R--) {$\Fock_R^{(-)}$};
		\node at (-0.5, 2.5) (L--) {$\Fock_L^{(-)}$};
		
		\node at (-1.5, -0.5) (R-+) {$\Fock_R^{(-)}$};
		\node at (-1.5, 1.5) (L-+) {$\Fock_L^{(-)}$};
		
		\node at (-0.5, -1.5) (R+-) {$\Fock_R^{(+)}$};
		\node at (1.5, -1.5) (L+-) {$\Fock_L^{(+)}$};
		
		\node at (2.5, 1.5) (R++) {$\Fock_R^{(+)}$};
		\node at (2.5, -0.5) (L++) {$\Fock_L^{(+)}$};
		
		\draw[->] (R--.west)+(0.2, 0) -- +(-1.2, 0);
		\draw[->] (R-+.north) -- +(0, 1.4);
		\draw[->] (R+-.east)+(-0.2, 0) -- +(1.2, 0);
		\draw[->] (R++.south) -- +(0, -1.4);
		
		\end{tikzpicture}
		\begin{tikzpicture}
		\draw[step=1cm,gray,very thin] (-2.9,-2.9) grid (3.9,3.9);
		\draw[pattern=north west lines, pattern color=black] (0.0,-2.9) rectangle (1.0,3.9);
		\draw[pattern=north east lines, pattern color=black] (-2.9,0.0) rectangle (3.9,1);
		
		\node at (1.5, 2.5) (R++) {$\Fock_R^{(+)}$};
		\node at (-0.5, 2.5) (L-+) {$\Fock_L^{(-)}$};
		
		\node at (-1.5, -0.5) (R+-) {$\Fock_R^{(+)}$};
		\node at (-1.5, 1.5) (L--) {$\Fock_L^{(-)}$};
		
		\node at (-0.5, -1.5) (R-+) {$\Fock_R^{(-)}$};
		\node at (1.5, -1.5) (L++) {$\Fock_L^{(+)}$};
		
		\node at (2.5, 1.5) (R--) {$\Fock_R^{(-)}$};
		\node at (2.5, -0.5) (L+-) {$\Fock_L^{(+)}$};
		
		\draw[->] (R--.west)+(0.2, 0) -- +(-3.2, 0);
		\draw[->] (R-+.north) -- +(0, 3.4);
		\draw[->] (R+-.east)+(-0.2, 0) -- +(3.2, 0);
		\draw[->] (R++.south) -- +(0, -3.4);
		\end{tikzpicture}
		\caption{Example locations of left and right modules within an arbitrary extended Kac table. Shaded regions indicate edge entries; their intersections are corner entries $(r, s) = (mp, nq)$ -- although we require principal diagonal corner entries ($m=n$) for the particular examples shown here. We have chosen to show $(N, \Delta h) = (1, -2)$ (left-hand diagram) and $(N, \Delta h)=(2, -2)$ (right-hand diagram). Horizontal arrows correspond to $\mu^-$, vertical ones to $\mu^+$. When $(p,q)$ are large enough compared to $(N,\Delta h)$, $\Fock_L$ and $\Fock_R$ will belong to the bulk (implicitly shown here), although in particular instances $\Fock_L$ and $\Fock_R$ may themselves be edge or even corner entries.\label{fig:ExtendedKacExample}}
	\end{figure}
	
	It is worth noting that some authors (e.g. \cite{MaRi08}) have considered `shifted' extended Kac tables, designed to contain entries at fractional $(r, s)$. Such modules were included in order to resolve some difficulties arising from the computation of fusion products. It is not clear (and remains to be studied) whether or not permitting fractional $(r, s)$ in the above corresponds also to these staggered modules.
	
	\subsection{Main Result}
	
	Equipped with this background knowledge, we now turn to the main result of this paper: a formula for the $\beta$-invariant of an arbitrary staggered Fock module of the type constructed in Section~\ref{sec:StaggeredFockModules}. We must assume that the singular vector at grade $h_R$ in the left-hand space $\Fock_L$ is the \textit{first} proper singular vector of that module. If this were not the case, then due to the submodule embedding structure of Fock space there would exist no $U_{\Delta h}$ which created $V_0\ket{\eta_R}$ from $\ket{\eta_L}$, and $\beta$ would be undefined. In any case, we can always project onto a subspace of $\Fock_L$ in which this holds, so the requirement is not a stringent one. We do not need to check that $V_0\ket{\eta_R}$ is actually a proper singular vector of $\Fock_L$. This follows trivially from the fact that $V_0$ is an intertwiner.
	
	Given spaces $\Fock_L$ and $\Fock_R$ and an intertwiner $V_0: \Fock_R \to \Fock_L$, the determination of $\beta$ then proceeds as follows:
	\begin{enumerate}
		\item Evaluate $V_0\ket{\eta_R}$ and calculate the constant $C_1$ such that $V_0 (C_1\ket{\eta_R}) = U_{\Delta h} \ket{\eta_L}$, where $U_{\Delta h} \in \univ{\Vir}_{\Delta h}$ is an operator such that 
		\begin{itemize}
			\item $U_{\Delta h} \ket{\eta_L}$ is (by assumption) the first proper singular vector of $\Fock_L$
			\item $U_{\Delta h}$ is normalised to have leading coefficient $U_{\Delta h} = L_{-1}^{-\Delta h} + \cdots$.
		\end{itemize}
		\item Evaluate $(U_{\Delta h})^\dagger \ket{\eta_R}$ in $\Stag$ and find $C_2$ such that $(U_{\Delta h})^\dagger \ket{\eta_R} = C_2 \ket{\eta_L}$.
		\item $\beta = \bra{\eta_L} (U_{\Delta h})^\dagger C_1\ket{\eta_R}$; in other words, $\beta = C_1C_2$.
	\end{enumerate} 
	
	We can use the facts derived in Section~\ref{sec:StaggeredFockModules} to compute $C_1$ and $C_2$ by simply comparing coefficients. In what follows, let $U_{\Delta h}$ be the following sum over partitions of $(-\Delta h)$:
	\begin{align*}
	U_{\Delta h} 		&= A_{(1^{-\Delta h})}L_{-1}^{-\Delta h} + A_{(2^11^{-\Delta h-2})}L_{-2}L_{-1}^{-\Delta h-2} + \cdots + A_{(-\Delta h^1)} L_{\Delta h} \\
	&=\sum_{|\tau| + \Delta h = 0} A_{\tau} L_{-(\tau)}.
	\end{align*}
	where $A_\tau \in \mathbb{C}$. 
	
	\subsubsection{Evaluation of $C_1$}
	
	We determine $C_1$ by comparing coefficients of $V_0\ket{\eta_R}$ and $U_{\Delta h}\ket{\eta_L}$ in the oscillator basis. For the latter, we must evaluate arbitrary (PBW-ordered) monomials of Virasoro modes -- a nontrivial task. This is made easier if we decide ahead of time which coefficient we would like to compare. There is an obvious choice; the term corresponding to the singleton $\alpha_{\Delta h}$. This is because many of the terms that appear in the oscillator expansion of a Virasoro monomial can be discarded -- once a product of $\alpha$s is produced, it can never again contribute to the singleton. 
	
	With this in mind, observe that
	\begin{align*}
	L_{-\tau_1} \cdots L_{-\tau_k} \ket{\eta} &= (\tau_k)(\tau_k + \tau_{k-1})\cdots (\tau_k + \cdots + \tau_2) \left[\eta + (\tau_k - 1)\lambda\right] \alpha_{-\sum_i \tau_i}\ket{\eta} \\
	& \quad + (\text{products of multiple } \alpha\text{s})\ket{\eta},
	\end{align*}
	therefore 
	\begin{equation}
	\label{eqn::U Delta h ket eta L}
	U_{\Delta h} \ket{\eta_L} = -\left(\sum_{|\tau| + \Delta h= 0}\frac{[\tau]!}{\Delta h}[\eta + (\tau_{\ell(\tau)} - 1)\lambda]A_{\tau}\right)\alpha_{\Delta h} \ket{\eta_L} + (\cdots).
	\end{equation}
	Since $U_{\Delta h} \ket{\eta_L}$ is the first proper singular vector of $\Fock_L$, it can be shown that this coefficient is never zero (Appendix~\ref{app:Non-Vanishing}). Of course, the summands of (\ref{eqn::U Delta h ket eta L}) themselves depend heavily on a choice of PBW-basis ordering. We choose the one which consists of all monomials of $\Vir$-generators with non-decreasing indexes -- i.e., $\tau$ is a partition. Then the notation $[\tau]!$ is intended to denote
	\[
	(\tau_1 + \tau_2 + \cdots + \tau_{\ell(\tau)})(\tau_2 + \cdots + \tau_{\ell(\tau)}) \cdots (\tau_{\ell(\tau)}), 
	\]
	a kind of `factorial' of the parts of $\tau$.
	
	We wish to compare this coefficient to the corresponding one in $V_0 \ket{\eta_R}$. We find
	\[
	V_0 \ket{\eta_R} = -\frac{\mu}{\Delta h} B^{(-\Delta h^1)}_N(\mu) \alpha_{\Delta h}\ket{\eta_L} + (\cdots)
	\]
	(recall the functions $B^{(\tau)}_N(\mu)$ were defined in (\ref{eqn:BmuIntegralFunction}). Therefore
	\begin{equation}
	\label{eqn:c1}
	C_1 = \frac{1}{\mu B^{(-\Delta h^1)}_N(\mu)}\left(\sum_{|\tau| + \Delta h= 0}[\tau]![\eta + (\tau_{\ell(\tau)} - 1)\lambda]A_{\tau}\right).
	\end{equation}
	$B^{(-\Delta h^1)}_N(\mu)$ is nonzero, otherwise $V_0 \ket{\eta_R}$ would not agree with $U_{\Delta h}\ket{\eta_L}$, which as discussed has a nonzero coefficient on the $\alpha_{\Delta h} \ket{\eta_L}$ term in its oscillator basis expansion.
	
	\subsubsection{Evaluation of $C_2$}
	
	This step proceeds very similarly to the above. Observe that in the full module $\Stag$,
	\begin{align*}
	L_{\tau_{k}} \cdots L_{\tau_1} \ket{\eta_R} &= (-\tau_1)(-\tau_1 -\tau_2) \cdots (-\tau_1 - \cdots -\tau_{k-1})V_{\sum_i \tau_i}\ket{\eta_R} \\
	&= \frac{[\overline{-\tau}]!}{-\sum_i \tau_i} \frac{1}{N\mu} [\alpha_{\sum_i \tau_i}, V_0]\ket{\eta_R} 
	\end{align*}
	where similarly to before, we write $[\overline{-\tau}]!$ to mean
	\[
	(-\tau_1 - \cdots - \tau_{\ell(\tau)})(-\tau_1 - \cdots - \tau_{\ell(\tau)-1}) \cdots (-\tau_1),
	\]
	a kind of `rising factorial' of the parts of a partition $\tau$, but with alternating sign.
	
	Now
	\begin{align*}
	(U_{\Delta h})^\dagger \ket{\eta_R} &= \left(\sum_{|\tau| + \Delta h= 0}\frac{[\overline{-\tau}]!}{\Delta h}\overline{A}_{\tau}\right) \frac{1}{N\mu}[\alpha_{-\Delta h}, V_0] \ket{\eta_R} \\
	&= \left(\sum_{|\tau| + \Delta h= 0}\frac{[\overline{-\tau}]!}{\Delta h}\overline{A}_{\tau}\right) \frac{1}{N\mu}[\alpha_{-\Delta h}, \frac{\mu}{-\Delta h} B^{(-\Delta h^1)}_N(\mu)\alpha_{\Delta h} + (\cdots)] \ket{\eta_L} \\
	&= \frac{B^{(-\Delta h^1)}_N(\mu)}{N}\left(\sum_{|\tau| + \Delta h= 0}\frac{[\overline{-\tau}]!}{\Delta h}\overline{A}_{\tau}\right) \ket{\eta_L},
	\end{align*}
	so
	\begin{equation}
	\label{eqn:c2}
	C_2 = \frac{B^{(-\Delta h^1)}_N(\mu)}{\Delta h N}\left(\sum_{|\tau| + \Delta h= 0}[\overline{-\tau}]!\overline{A}_{\tau}\right)
	\end{equation}
	
	\subsubsection{Theorem and Examples}
	\label{subsubsec:TheoremAndExamples}
	
	We have proven:
	\begin{theorem*}[Main Result] 
		\label{thm::MainResult}
		Let $\Fock_L$ and $\Fock_R$ be two Fock spaces which are related by the intertwiner $V_0$. Suppose $V_0$ arises as the zero mode from the composition of $N$ many vertex operators. Let the operator which creates the first proper singular vector $U_{\Delta h}\ket{\eta_L}$ of $\Fock_L$ have the following explicit form:
		\begin{align*}
		U_{\Delta h} 		&= A_{(1^{-\Delta h})}L_{-1}^{-\Delta h} + A_{(2^11^{-\Delta h-2})}L_{-2}L_{-1}^{-\Delta h-2} + \cdots + A_{(-\Delta h^1)} L_{\Delta h} \\
					&=\sum_{|\tau| +\Delta h = 0} A_{\tau} L_{-(\tau)},
		\end{align*}
		and suppose $V_0 \ket{\eta_R} \propto U_{\Delta h} \ket{\eta_L}$. Then there exists a staggered Fock module $\Stag$ which fits into the short exact sequence
		\[
		0 \longrightarrow \Fock_{L} \longrightarrow \Stag \longrightarrow \Fock_{R} \longrightarrow 0
		\] 
		with
		\begin{equation}
		\label{eqn::MainResult}
		\beta(\Stag) = \frac{1}{\Delta h N \mu^2} \left(\sum_{|\tau| + \Delta h = 0}\left[\frac{1}{2}(N+\tau_{\ell(\tau)})\mu^2  - \tau_{\ell(\tau)} + \frac{\Delta h }{N}\right][\tau]! A_{\tau}\right)\left(\sum_{|\tau| + \Delta h = 0} [\overline{-\tau}]! \overline{A}_\tau\right).
		\end{equation}
	\end{theorem*}

	Notationally, we have left the normalisation of $U_{\Delta h}$ unspecified. In practice, we choose to set $A_{(1^{-\Delta h})} = 1$, but changing convention is merely a matter of scaling by $|A_\tau|^2$ for the appropriate $A_\tau$. 
	
	We now apply these results to particular instances of $(N, \Delta h)$, expressing $\beta$ as a function of $\mu$. This is possible as the $A_\tau$ may be written in terms of $c$ and $h_L$, which themselves have expressions in terms of $\mu$. Since the degree of the first proper singular vector of $\Fock_L$ is only dependent on $\Delta h$, it is convenient when deriving general formulae to work with $N$ initially unspecified. The $A_\tau$ only depend on $N$ implicitly, through $h_L$, so need only be calculated once for each $\Delta h$. Because of this, large families of $\beta(\mu)$ can essentially be derived simultaneously. Figure~\ref{fig:betaFunctions} contains some explicit examples, but we present here the formulae at arbitrary $N$ for $\Delta h = -1, -2, -3$:
	\begin{align*}
	\left.\beta(\mu, N)\right|_{\Delta h = -1} 	&= \frac{N + 1}{2N^2}\mu^{-2}(N\mu^2 - 2)\\ 
	\left.\beta(\mu, N)\right|_{\Delta h = -2} 	&= \frac{(N + 1)(N + 2)}{36N^4}\mu^{-2}\left(N^2(N^2 -1)\mu^4 + 6N^2\mu^2 + 16 - 4N^2\right)\\ \quad 		
	&\times\left(N(N-1)\mu^2 + 2N - 4\right)(N\mu^2 - 2)(N\mu^2 - 4)\\ 
	\left.\beta(\mu, N)\right|_{\Delta h = -3} &= \frac{3}{2048N^{10}}\mu^{-10}(N(N+1)\mu^2 + 2N + 6)(N(N+1)\mu^2 - 2N + 6)\\ &\times(N(N+1)\mu^2 - 2N - 6)^2(N(N-1)\mu^2 + 2N + 6)\\ &\times(N(N-1)\mu^2 + 2N - 6)^2(N(N-1)\mu^2 - 2N + 6)(N^2\mu^2 - 6).
	\end{align*}
	As can be seen, these expressions quickly become very complicated, without obvious pattern. One can observe from the explicit examples in the table that whenever $\frac{\Delta h}{N} \in \mathbb{Z}$, there appear to be several collisions of the factors involved, so that the entire expression becomes much simpler.
	
	These formulae hold for generic $\mu$, but are of most interest when $\mu = \mu^\pm$ of a reducible $(p, q)$ theory. We note that when we substitute particular $(p, q)$ into these equations, we reproduce the results of various authors (Figure~\ref{fig:betaCalculationsEta-Mu-}). This is suggestive that Fock spaces provide effective means of computation for staggered modules, and that modes of vertex operators are the `correct' staggering operators for free field realisations of them. Particular values of interest from the literature are $(N, \Delta h) = (2, -2)$ and $(1, -2)$ because at $(p, q) = (2, 3)$ the former should correspond to critical percolation with $\beta = \frac{-5}{8}$ and the latter to the dilute polymers with $\beta = \frac{5}{6}$. Accounting for normalisation (a factor of $\frac{q^2}{p^2}$ for $(N, \Delta h) = (2, -2)$ and $\frac{p^2}{q^2}$ for $(1, -2)$), we find this to be the case.
	
	Some authors derive or suggest general formulae for $\beta$ in terms of $p$ and $q$ for particular types of staggered module (e.g. the $LM(2,q)$ modules in \cite{MaRi08}). These are special cases of the formula (\ref{eqn::MainResult}). In some cases, these authors have commented on the surprisingly neat way in which $\beta(t)$ splits into linear factors, where $t$ parametrises the central charge $c$ as
	\[
	c = 13 - 6(t + \frac{1}{t}).
	\]
	Clearly $t = \mu^2$. We can immediately deduce that only even powers of $\mu$ may appear, as not only is every explicit appearance of $\mu$ in (\ref{eqn::MainResult}) of the form $\mu^{2n}$, but also every implicit appearance via the $A_\tau$: they involve only integral powers of $h_L$ and $c$, with rational coefficients.
	
	The fact that $\beta$ factorises so neatly may be related to the structure of the $(p, q)$ extended Kac tables. Depending on the values of $p$ and $q$, the entries related via staggered module with some fixed $(N, \Delta h)$ may belong to the bulk, to the edges, or even to the corners. Of course, corner Fock modules have \textit{no} submodule embedding structure, and are instead completely reducible (ref. Figure~\ref{fig:FockEmbedding}). This means that $\beta$ is undefined, since there is no operator $U_{\Delta h}$ which creates the first proper singular vector. Of course, since the formulae for $\beta$ are just polynomials, we can still evaluate them at these points.
	
	When $\frac{\Delta h}{N} \in \mathbb{Z}$, this `corner collision' occurs whenever $N$ is a multiple of $p$ and $\frac{\Delta h}{N}$ a multiple of $q$, or vice versa, according to $\mu = \mu^+$ or $\mu^-$ respectively. Therefore, $\beta(\mu)$ should be undefined at the collection of points
	\[
	\prod_{p|N, \; q|\frac{\Delta h}{N}, \; \gcd(p, q) = 1}\left(\mu - \sqrt{\frac{2q}{p}}\right), \qquad \prod_{q|N, \; p|\frac{\Delta h}{N}, \; \gcd(p, q) = 1}\left(\mu + \sqrt{\frac{2p}{q}}\right)
	\]
	from these two contributions. By noting symmetry, we see that this is actually a net factor of
	\[
	\prod_{p|N, \; q|\frac{\Delta h}{N}, \; \gcd(p, q) = 1}\left(\mu^2 - \frac{2q}{p}\right) 
	\]
	When applied to the formulae seen in Figure~\ref{fig:betaFunctions}, we see an interesting empirical relationship between these special values and the zeroes of $\beta$, in that we correctly predict some of the factors in $\mu^2$ -- but not all. As well as missing some of the type $(\mu^2 - a)$, those of the form $(\mu^2 + a)$ are entirely absent, which we note \textit{cannot} come from this kind of consideration of the extended $(p, q)$ Kac tables with $p, q > 0$. This perhaps hints that the $\beta$ invariant `knows about' other $c > 1$ staggered $\Vir$-modules arising from generic integers $(p, q)$.
	
	
		\begin{landscape}
		\begin{figure}
			\centering
			\[
			\begin{array}{|c|c|c|c|}
			\hline
			N\backslash \Delta h & -1 & -2 & -3 \\\hline
			1 & \mu^{-2}(\mu^2 - 2 ) & -2\mu^{-2}(\mu^4 - 4)(\mu^2 - 4) & 12\mu^{-10}(\mu^4-16)(\mu^4 - 4)(\mu^2 - 6)\\ \hline
			2 & \frac{3}{4}\mu^{-2}(\mu^2 -1) & \frac{1}{2}\mu^{2}(\mu^4-4)(\mu^2-1) & \begin{array}{c}\frac{45}{2^{14}}\mu^{-10}(9 \mu^4-25)(\mu^4-1)(\mu^2-3)\\ \times(\mu^2-1)^2(3\mu^2+1)(\mu^2+5) \\ \end{array} \\ \hline
			3 & \frac{1}{3}\mu^{-2}(3\mu^2 - 2) & \frac{20}{729}\mu^{-2}(36\mu^4 + 27\mu^2 - 10)(3\mu^2 + 1)(3\mu^2 - 2)(3\mu^2 - 4) & \mu^{-2}(\mu^4 - 4)(\mu^4-1)(3\mu^2-2)\\ \hline
			4 & \frac{5}{16}\mu^{-2}(2\mu^2 - 1) & 5\mu^{-2}(5\mu^4 + 2\mu^2 - 1)(3\mu^2 + 1)(2\mu^2 - 1)(\mu^2 - 1) & \begin{array}{c}\frac{3}{4194304}\mu^{-10}(10\mu^2 + 7)(10\mu^2 - 1)(10\mu^2 - 7)^2\\ \times(8\mu^2 - 3)(6\mu^2 + 7)(6\mu^2 + 1)^2(6\mu^2 - 1) \end{array}\\ \hline
			5 & \frac{3}{25}\mu^{-2}(5\mu^2 - 2) & \frac{14}{625}\mu^{-2}(100\mu^4 + 25\mu^2 - 14)(10\mu^2 + 3)(5\mu^2 - 2)(5\mu^2 - 4) & \begin{array}{c} \frac{6}{9765625}\mu^{-10}(25\mu^2 - 6)(15\mu^2 + 8)(15\mu^2 - 2)\\ \times(15\mu^2 - 8)^2(5\mu^2 + 4)(5\mu^2 + 1)^2(5\mu^2 - 1) \end{array} \\ \hline
			6 & \frac{7}{36}\mu^{-2}(3\mu^2 - 1) & \frac{28}{729}\mu^{-2}(315\mu^4 + 54\mu^2 - 32)(15\mu^2 + 4)(3\mu^2 - 1)(3\mu^2 - 2) & \begin{array}{c}\frac{1}{4096}\mu^{-10}(7\mu^2 + 3)(7\mu^2 - 1)(7\mu^2 - 3)^2\\ \times(6\mu^2 - 1)(5\mu^2 + 3)(5\mu^2 + 1)^2(5\mu^2 - 1)\end{array}\\ \hline
			\end{array}
			\]
			\caption{Various $\beta(\mu)$ at low-lying $(N,\Delta h)$. Be aware that some expressions my only apply for certain $(p, q)$, as per the discussion in Section~\ref{subsec:AdmissibleLandR}.\label{fig:betaFunctions}}
		\end{figure}
	\end{landscape}
	
	
	\begin{landscape}
		\begin{figure}
			\renewcommand{\arraystretch}{2.5}
			\centering
			\begin{tabular}{|c|c|c||ccccccc|c|} \hline
				\textbf{Dense logCFTs}& $(p, q)$ & $c$ & $\beta_{1, 3}$ & $\beta_{1, 4}$ & $\beta_{1, 5}$ & $\beta_{1, 6}$ & $\beta_{1, 7}$ & $\beta_{1, 8}$ & $\beta_{1, 9}$ & \textbf{Ref} \\ \hline
				- & $(1, 4)$ & $-\frac{25}{2}$& & & & & & & $\underset{(1,-1)}{-3}$ & \cite{VJS11} \\
				-& $(1, 3)$ & $-7$&	& & & & $\underset{(1,-1)}{-2}$ & $\underset{(2,-2)}{8}$ & & \cite{VJS11} \\
				Dense Polymer & $(1, 2)$ & $-2$& & & $\underset{(1,-1)}{-1}$ & & $\underset{(1,-2)}{-\frac{9}{2}}$ & & $\underset{(1,-3)}{-\frac{75}{4}}$ & \cite{Car13, Gur13, VJS11} \\
				Percolation & $(2, 3)$ & $0$& & $\underset{(1,-1)}{-\frac{1}{2}}$ & $\underset{(2,-2)}{-\frac{5}{8}}$ & &  $\underset{(1,-3)}{-\frac{35}{3}}$ & $\underset{(2,-6)}{-\frac{13475}{216}}$ & & \cite{Car13, Gur13, VJS11}\\
				Ising&$(3, 4)$&$\frac{1}{2}$&&&$\underset{(1,-2)}{-\frac{35}{24}}$&$\underset{(2,-4)}{-\frac{13475}{243}}$&$\underset{(3,-6)}{-\frac{49049}{17496}}$&&$\underset{(1,-5)}{-\frac{40415375}{944784}}$& \cite{VJS11}\\
				Tricritical Ising& $(4, 5)$&$\frac{7}{10}$&&&&$\underset{(1,-3)}{-\frac{693}{100}}$&$\underset{(2,-6)}{-\frac{6114399291}{1078465600}}$&$-\frac{3^{14}7^411^213^217^123^1}{2^{12}5^219^111423^2}$(?)&$-\frac{3^{13}7^213^317^219^123^1}{2^65^411^218433^2}$(?)& \cite{VJS11}\\
				3-State Potts& $(5, 6)$&$\frac{4}{5}$&&&&&$\underset{(1,-4)}{-\frac{676039}{59895}}$&$-\frac{2^{12}7^211^317^219^123^229^1}{3^25^341^217234^2}$(?)&$-\frac{3^97^611^313^217^319^123^229^1}{5^59887^2281623^2}$(?)&\cite{VJS11} \\\hline \hline
				\textbf{Dilute logCFTs}& $(p, q)$ & $c$ & $\beta_{3, 1}$ & $\beta_{4, 1}$ & $\beta_{5, 1}$ & $\beta_{6, 1}$ & $\beta_{7, 1}$ & $\beta_{8, 1}$ & $\beta_{9, 1}$ & \textbf{Ref} \\ \hline
				Dilute Polymer & $(2, 3)$ & $0$& $\underset{(1,-2)}{\frac{5}{6}}$ &  & $\underset{(1,-5)}{\frac{67375}{676}}$ & &  $\frac{2^15^57^211^213^217}{3^123^2271^2}$(?) & & $\frac{5^67^511^313^21^219^223^1}{2^43^123607803^2}(?)$ & \cite{VJS11}\\
				$O(n\to1)$ Ising&$(3, 4)$&$\frac{1}{2}$&&$\underset{(1,-3)}{\frac{175}{12}}$&$\underset{(2,-6)}{\frac{49049}{15552}}$&&$\frac{5^27^111^213^117^119^1}{2^13^2113^2}$(?)&$\frac{5^97^511^513^417^219^223^1}{2^{12}29^247^2787^21201^2}$(?)&& \cite{VJS11} \\ \hline
			\end{tabular}
			\caption{Known values of $\beta$ for various staggered modules. The $A_{(-\Delta h^1)}=1$ normalisation convention is used, not $A_{(1^{-\Delta h})} = 1$, to match with those in the referred works.  Unless marked with a (?), entries in the table have been verified to match with those predicted by (\ref{eqn::MainResult}) -- i.e., these staggered modules do indeed have Fock space resolutions of the type constructed here. The relevant $(N, \Delta h)$ are given in small font beneath each value. Blank spaces indicate trivial or absent staggering. $\beta_{r, s}$ indicates that the right module occurs at $(r, s)$ in the appropriate extended table. Dense logCFTs take $\mu = \mu^-$, dilute ones $\mu = \mu^+$, and both have $\eta = \eta^-$. There is then only one way to assign the left module, at least for the small values of $(r,s)$ considered here.} \label{fig:betaCalculationsEta-Mu-} 
		\end{figure}
	\end{landscape}

%
	
	\subsection{Degenerate Staggerings and Algebraic Construction of $\Stag$}
	\label{subsec::algConstruction}
	
	Using our construction, no staggered Fock module $\Stag$ may exist with $\Fock_L \cong \Fock_R$. This would correspond to $\mu = 0$, where all vertex operators are trivial. We see this as a pole appearing in each $\beta(\mu)$.
	
	However, notice that
	\[
	[L_m, \alpha_n - \lambda\delta_{n, 0}] - [L_n, \alpha_m - \lambda\delta_{m, 0}] = (m-n)(\alpha_{m+n} - \lambda\delta_{m+n, 0}),
	\]
	so in fact 
	\[
	\{ \alpha_n - \lambda \delta_{n, 0} | n\in\mathbb{Z}\}
	\]
	is a valid set of staggering operators, this time for the sequence
	\[
	0 \longrightarrow \Fock \longrightarrow\Stag\longrightarrow\Fock\longrightarrow 0,
	\]
	for \textit{any} Fock space $\Fock$! Of course any $\beta$-invariant for such a module is trivial. This does not, however, mean that the staggering is trivial. This family of staggering operators is distinctive in that the $\Vir$-module homomorphisms in the short exact sequence \textit{do} descend to obvious $\Heis$-module homomorphisms. This can be seen by defining the $\Heis$ action on $\Stag$ to be
	\[
	\alpha_n = \matII{\alpha_n}{\delta_{n, 0}}{0}{\alpha_n}.
	\]
	Then one may easily verify that
	\[
	L_n = \frac{1}{2}\sum_{k \in \mathbb{Z}} \normord{\alpha_k \alpha_{n-k}} - (n+1)\lambda \alpha_n
	\]
	still holds true in $\Stag$.
	
	The logarithmic behaviour in this example is due to the presence of the field 
	\[
	a(z) = q + \ln(z)\alpha_0 - \sum_{n\neq0} \frac{\alpha_n}{n}z^{-n}, \qquad a(z)a(w) \sim \ln(z-w)
	\]
	(which, we pause to mention, can be seen as the linearisation of $V_\mu(z)$ in the limit $\mu\to0$). $a(z)$ is not typically included in the VOA as it is actually $\partial a(z)$ and its derivatives which generate the space of states, but 
	\begin{align*}
	a(z)\ket{0} &\sim q \ket{0}\\
	a(z) \ket{\eta} &= a(z)e^{\eta q}\ket{0} \\ &\sim \ln(z)\alpha_0 \ket{\eta} + q \ket{\eta} 
	\end{align*}
	which is suspiciously like the results seen in e.g. \cite{Gur13}, where a vacuum vector is created alongside its logarithmic partner, a $\ln(z)$ factor appearing on the former.
	
	Does the field $a(z)$ indeed create the logarithmic partner state? Well, since $[\alpha_0, q] = 1$, we find that
	\[
	[L_n, q] = \alpha_n - \lambda\delta_{n, 0}
	\]
	reproduces the staggering operators. For these modules, then, we have the following algebraically-minded construction:
	\[
	\Stag \cong \frac{\mathbb{C}[q]}{\left<q^2\right>}\otimes\Fock_\eta
	\]
	where $q\ket{\eta}$ is the logarithmic partner state of the vacuum $\ket{\eta}$, since
	\[
	L_n q\ket{\eta} = qL_n\ket{\eta} + (\alpha_n - \lambda\delta_{n,0})\ket{\eta}
	\]
	recreates the $\Vir$-action on $\Stag$. Compare this to the induced module construction of a Fock space, where we extend a $1$-d representation of the subalgebra of non-negative modes to the entirety of $\univ{\Heis}$. If $q$ is included into this subalgebra, then we recover standard Fock space by also letting it act trivially on the vacuum. $\Stag$ is the module induced by instead allowing $q^2$ to act trivially. Larger (i.e. higher rank) staggered modules of this type may of course be considered, simply by replacing the $\left<q^2\right>$ quotient by some higher-order expression in $q$. Note the similarities between this construction and that of \cite{Flo98}, where expansion in a nilpotent variable $\theta^r = 0$ allowed for much simpler expressions for generating fields and their differential relationships within rank $r$, $\Delta h = 0$ staggered modules. Similar ideas have appeared more recently in e.g. \cite{HPV16}. We therefore speculatively suggest that it is the inclusion of $a(z)$ in the field-theoretic data that naturally permits the existence of this type of staggered Fock module. 
	
	Where does this result take us when we turn to the more general case, with $\Fock_L \neq \Fock_R$? Can we find a similar algebraic construction of these $\Stag$? We would need to find or construct (or define) an operator $Q$ which satisfied $[L_n, Q] = V_n = \frac{1}{N\mu}[\alpha_n, V_0]$. Given such a $Q$, we would then have
	\[
	\Stag \cong \frac{\mathbb{C}[Q]}{\langle Q^2 \rangle} \otimes \Fock_R
	\]
	for the staggered Fock module arising from the vertex operator $V_{N\times\mu}(z)$. Working with the following expansion of (\ref{eqn:NormOrdNMany}) in the variables (\ref{eqn:ProjectiveVariables}) and collecting all terms of fixed degree into coefficients $\left[V_n(y)\right]$,
	\[
	V_{N\times\mu}(x; y) = \sum_{n \in \mathbb{Z}} \left[V_n(y)\right] x^{-n-1}.
	\]
	If we write
	\[
	W_{N\times\mu}(x) = Q + \ln(x)\left[V_0(y)\right] - \sum_{n \neq 0}\frac{1}{n}\left[V_n(y)\right] x^{-n},
	\]
	then similarly to before (at least for the $c\leq1$ Fock spaces, where $\mu \in \mathbb{R}$),
	\begin{align*}
	W_{N\times\mu}(x; y)\ket{0} &\sim \lim\limits_{x \to 0}W_{N\times\mu}(x; y)\ket{0}\\
	&= Q\ket{0},\\
	W_{N\times\mu}(x; y) \ket{\eta_R} &\sim \ln(x)\left[V_0(y)\right]\ket{\eta_R} + Q\ket{\eta_R} 
	\end{align*}
	where one must remember that the degrees of the operators $\left[V_n(y)\right]$ depend upon the value of $\eta$, so the zero -indexed mode is a different operator in the two cases. We also have $V_{N\times\mu} = \partial_x W_{N \times \mu}$.
	
	The staggering operators $\{\frac{1}{N\mu}[\alpha_n, V_0] | n\in\mathbb{Z}\}$ seem to arise from the expansion $\partial a(z) V_{N\times \mu}(w)$. We instead suggest that we should be looking at $a(z)V_{N\times \mu}(w)$, ignoring for the moment the fact that logarithmic terms arising here damage our ability to interpret contour integrals. This means, algebraically, we should consider the object
	\[
	Q = \frac{1}{N\mu}[q, V_0].
	\]
	We find by application of the Jacobi identity that
	\[
	[L_n, Q] = V_n,
	\]
	therefore this object $Q$ interacts with the Virasoro algebra in the desired way. However, unlike before, there is no comparable induced module construction of $\Stag$ -- because one does not typically generate a representation space from the modes $V_n$ of a single screening operator. However, there is in principle no reason why this cannot be done. Such a thing may be necessary in developing the theory of logarithmic vertex operator algebras, where the state spaces are the staggered modules. To perform a more careful analysis of this would go beyond the scope of this current work, so we defer going into details.
	
	\section{Concluding Remarks}
	\label{sec::conclusion}

	It appears that the application of Fock spaces to the construction of staggered Virasoro modules is very successful, giving an explicit formula for $\beta$ which agrees with many known cases. One (previously unmentioned) clue that this should be the case is in the submodule embedding structure (Figure~\ref{fig:FockEmbedding}). Whereas previous authors have had to make (arguably) ad-hoc module constructions from quotients of Verma modules in order to achieve left and right modules with singular vectors at the correct conformal weights, Fock spaces instead allow us to interchange the two vacuuum momenta $\eta^\pm$ to actively select which of two possible choices is to be the first proper singular vector. Then, the entire staggered module $\Stag$ is generated by the action of $\Vir$ on the vacuum vector of the right-hand space, $\ket{\eta_R}$ (when we consider $\FockS_L \oplus \FockS_R$ as the staggered module).
	
	Additionally, the structure of the Kac tables is suggestive of a simple explanation for some, but not all, of the zeroes of $\beta$. This result remains conjectural. One logical conclusion of this idea is that there may exist interesting staggered structures with `natural' interpretations within $(p, q)$ Kac tables for negative $p, q$. 
	
	As noted above, whether or not the object $W_{N\times \mu}(z)$ is meaningful beyond the algebraic construction is unclear. A more careful analysis of the ramifications of including it in the field theory needs to be done before any such claims can be made. We might expect to find, like with the inclusion of $a(z)$, that the inclusion of a single weight-$0$ permits the existence of staggered modules of a particular type, of arbitrarily high rank (through OPEs of this field).
	
	Finally, it should be mentioned that the use of screening operators is certainly not a requirement for the existence of staggered Fock modules, and that for the construction utilised here, \textit{any} valid family of staggering operators may be used. 
	
	\subsection{Acknowledgements}	
	
	The author would like to acknowledge the valuable contributions and suggestions made by Peter Bouwknegt and by David Ridout, and the generous support given by the School of Mathematics and Physics at the University of Tasmania for portions of this research. 
	
	\appendix
	
	\section{Generalised Selberg Integrals}
	\label{app:Selberg}
	
	We previously wrote 
	\[
	B^{(\tau)}_N(\mu) = \oint_\mathcal{S} \prod_{\tau_j \in \tau} \left(1 + p_{\tau_j}(y)\right)\prod_{i < j} (y_i - y_j)^{\mu^2} \prod_i (1- y_i)^{\mu^2} y_i^{\mu\eta_R} \de y_i,
	\]
	but were able to avoid its evaluation by using the insensitivity of $V_0$ to overall scaling. Now we use the same kind of technique to assign values to such integrals, up to normalisation by $B^{(n^1)}_N(\mu)$.
	
	Firstly suppose $U_{\Delta h}\ket{\eta_L}$ has expansion in the oscillator basis
	\begin{align*}
	U_{\Delta h} \ket{\eta_L} 	&= C_{(1^{-\Delta h})} \alpha_{-1}^{-\Delta h} + \cdots + C_{(-\Delta h ^1)} \alpha_{\Delta h} \\
		&= \sum_{|\tau| + \Delta h = 0} C_\tau \alpha_{-(\tau)}.
	\end{align*}
	Then we compute $\beta$ by comparing $U_{\Delta h} \ket{\eta_L}$ and $V_0 \ket{\eta_R}$ at this coefficient;
	\[
	\beta = \frac{C_\tau}{\Delta h N} \left( \prod_{i = 1}^{\infty} n_\tau(i)!\right)\left(\prod_{\tau_i\in \tau}\frac{\tau_i}{\mu}\right)\left(\sum_{|\sigma| + \Delta h = 0} [\overline{-\sigma}]! \overline{A}_\sigma\right)\frac{B^{(-\Delta h ^1)}_N(\mu)}{B^{(\tau)}_N(\mu)},
	\]
	and therefore find
	\[
	\frac{B^{(\tau)}_N(\mu)}{B^{(n^1)}_N(\mu)} = \frac{C_\tau \left( \prod_{i = 1}^{\infty} n_\tau(i)!\right)\left(\prod_{\tau_i \in \tau}\tau_i\right)}{\mu^{\ell(\tau) + 2} \left( \sum_{|\sigma| = n} \left[ \frac{1}{2}(N+\sigma_{\ell(\sigma)})\mu^2 - \sigma_{\ell(\sigma)} - \frac{n}{N} \right] [\sigma]! A_\sigma\right)}
	\]
	whenever there exists a staggering with the required parameters.
	
	\section{Non-Vanishing Nature of Singular Vector Coefficients}
	\label{app:Non-Vanishing}
	
	Let $\kappa: \Fock_\eta \to \mathbb{Z}$ be defined by
	\[
	\kappa(v) = \max\{k \in \mathbb{Z}| \alpha_k v \neq 0\}.
	\]
	
	Note that $\kappa(v) > 0$ for every $v\in \Fock_\eta$ except for $\ket{\eta}$. Now suppose that $v$ is actually a $\Vir$-singular vector. It is very easy to show that $\alpha_{\kappa(v)} v$ must also be singular, and cannot vanish (by assumption). Therefore the first proper singular vector of any Fock space must have a nonzero projection onto $\mathbb{C}\alpha_{\deg(v)}$.
	
	What is more, we may also easily show that
	\[
	\univ{\Vir} \cdot \alpha_k v \supseteq \univ{\Vir} \cdot \alpha_{\kappa(v)} v \quad \forall k
	\]
	which heavily restricts the embedding structure of the `degenerate' $(\Fock_L = \Fock_R)$ staggered modules discussed above.
	
	Since Fock space singular vectors are related to rectangular-partition Jack polynomials by a simple algebra homomorphism, this gives us a differential relationship between these polynomials. That is, for every such polynomial $J_{(m^n)}(x_1, x_2, \ldots)$ with highest-indexed variable $x_k$, 
	\[
	\frac{\partial}{\partial x_k} J_{(m^n)} \propto J_{(i^j)}
	\]
	for some $J_{(i^j)}(x_1, x_2, \ldots)$ such that both polynomials correspond to singular vectors in the same Fock space.
	
	\bibliographystyle{abbrv}		
	\bibliography{paper16bib}
	
\end{document}